\documentclass[%
 reprint,
 amsmath,amssymb,
 aps,
]{revtex4-1}

\usepackage{graphicx}% Include figure files
\usepackage{dcolumn}% Align table columns on decimal point
\usepackage{subscript}
\usepackage{bm}% bold math
\usepackage{hyperref}% add hypertext capabilities
\begin{document}

%\preprint{APS/123-QED}

\title{Metal-Insulator Transition in Ga doped ZnO via Controlled Thickness}
\author{Joynarayan Mukherjee}
\affiliation{Nano Functional Materials Technology Centre, Department of Physics,
Indian Institute of Technology Madras, Chennai, India, 600036}
\author{B. R. K. Nanda}
\email{nandab@iitm.ac.in}
\affiliation{Condensed Matter Theory and Computational Lab, Department of Physics,
Indian Institute of Technology Madras, Chennai, India, 600036}
\author{M. S. Ramachandra Rao}
\email{msrrao@iitm.ac.in}
\affiliation{Nano Functional Materials Technology Centre, Department of Physics,
Indian Institute of Technology Madras, Chennai, India, 600036}
\date{\today}% It is always \today, today,

\begin{abstract}
We report thickness dependent metal insulator transition in Ga doped ZnO (Ga:ZnO) thin films grown by pulsed laser deposition technique. From the electrical transport measurements, we find that
while the thinnest film exhibits a resistivity of 0.05 $\Omega$-cm,
lying in the insulating regime, the thickest has resistivity of $6.6\times10^{-4}\Omega$-cm
which shows metallic type of conduction. Our analysis reveals that the Mott's variable range hopping (VRH)
model governs the insulating behavior in the thinner Ga:ZnO whereas
the 2D weak localization phenomena is appropriate to explain the electron transport
in the thicker Ga:ZnO. Magnetoresistance study further confirms the presence of strong localization in 6 nm film while weak localization is observed in 20 nm and above thicker films. From the density functional calculations, it is found that due to surface reconstruction and Ga doping, strong crystalline disorder sets in very thin films to introduce localized states and thereby, restricts the donor electron mobility.
\end{abstract}
\maketitle

\section{Introduction}

For last many decades, ZnO has remained a material of
great research interest both in fundamental and applied aspects. Due
to high excitonic binding energy ($\thicksim$60 meV) and wide
bandgap of 3.37 eV, ZnO has potential applications in optoelectronics
devices like light emitting diodes (LED), photodetectors and laser
diodes\cite{Chu2008_ZnO_laser_diode_app, Choi2010_ZnO_LED_app, Chu2011_ZnO_lasing}. In particular, this compound has been a subject of
active investigation to exploit its potential as a transparent conducting
oxide which can serve as an alternate to indium tin oxide (ITO)\cite{Minami_2000_TCO}. The conductivity in n-type doped ZnO can be enhanced
as high as $\thicksim$7000 S/cm through substitution of group
III and IV elements at the Zn site\cite{Bhosle2006_MIT_Ga_conc, Kuznetsov2015_SiZnO_conductivity, Das2014_SiZnO_MIT}. The large conductivity in this system is primarily attributed to the delocalized p-obitals carrying
the donor electrons\cite{Sans2009_doped_ZnO_BS}. Among these n-type dopants, Ga is
the most suitable candidate to enhance the conductivity of ZnO since
both Ga and Zn have almost similar ionic radii which helps in increasing
the solubility of a dopant in the host material.

Recently, Look \textit{et al}. has achieved a minimum resistivity of $1.96\times10^{-4}$
$\Omega$-cm with 3\% Ga doping in ZnO thin films grown by pulsed
laser deposition (PLD) technique\cite{Look2011_GaZnO_res_mobility}. Earlier, Bhosle
\textit{et al}. had reported that 5\% Ga doping yields minimum resistivity
in ZnO thin films\citep{Bhosle2006_MIT_Ga_conc}. It is observed that resistivity first
decreases as doping concentration increases and after reaching a minimum
value, it again increases since dopant itself starts acting as a scattering
center\cite{Bhosle2006_MIT_Ga_conc}. Therefore, ZnO can exhibit metal insulator transition
(MIT) by tuning the factors such as doping concentration. In fact,
MIT in Ga doped ZnO as a function of doping concentration has been reported by several other groups \cite{Bhosle2006_MIT_Ga_conc, Snure2007_MIT_Ga_conc_Mott_limit}. With sufficiently high doping
concentration, the dopant electrons make ZnO degenerate as these electrons form a dispersive band. The MIT can
also be achieved in degenerately doped ZnO by reducing the thickness. It has been experimentally observed that the conducting
Si doped ZnO becomes highly resistive below a critical thickness of
15 nm\cite{Das2014_SiZnO_MIT}. However, so far the thickness dependent MIT investigation is
limited to Si doping only. Therefore, a detailed mechanism that causes
the MIT has not been completely understood yet. In this work, we examine the possibility of MIT in Ga:ZnO, which shows maximum conductivity among all n-type dopants, as a function of film thickness.

To gain an in-depth understanding behind the thickness driven MIT,
experimentally it is imperative to study the low temperature electrical
properties of degenerately Ga:ZnO films as a function of thickness.
Temperature dependent resistivity measurement undoubtedly reveals
the metallic or insulating nature. For metallic sample temperature
coefficient of resistance (TCR) is positive whereas for insulator
TCR becomes negative. In addition, the occupancy and k-dispersion
of the bands are good measures to examine the metallicity and thereby,
the MIT. In this context, solutions to the quantum mechanical many
body problems through \textit{ab initio} calculations based on density-functional
theory (DFT) are desirable.

In this paper, we present the structural, electrical and magnetotransport
properties of Ga:ZnO thin films of varying thickness, in the range
of 6 \textendash{} 51 nm, grown on c-sapphire substrate. From the
temperature dependent resistivity measurement, we observe MIT in Ga:ZnO films as a function of film thickness. For 6 nm thick
film, the resistivity is found to be close to 0.05 $\Omega$-cm which
lies in the insulating regime and conduction is governed by the variable range hopping (VRH) mechanism. For the film with thickness 20 nm and above,
the resistivity is of the order of $10^{-3}$ $\Omega$-cm. These films exhibit 2D weak localization phenomena at low temperature regime. Temperature dependent resistivity and magnetoresistance data confirm the presence of weak localization in films with thickness 20 nm and more while in 6 nm thick film exhibits strong localization behavior. Our density functional calculations suggest that, in thicker films, lattice distortion due to Ga dopant is weak and donor p-electron partially occupies a dispersive band in k\textsubscript{x}-k\textsubscript{y} plane which is the source of metallic type of conduction. However, in 6 nm film, surface reconstruction and the dopant induce sufficient lattice disorders which in turn create localized states in the vicinity of Fermi level and hence, the film exhibits insulating behavior.
\section{Experimental details}

1\% Ga doped ZnO (Ga:ZnO) thin films were deposited on (001) sapphire substrate
(5 mm $\times$ 5 mm) by PLD technique.
A KrF laser of wavelength 248 nm of fluence 2 J/cm\textsuperscript{2}
and 2 Hz pulse frequency was used for the deposition. The substrate
to target distance was 4.5 cm. During the deposition, O\textsubscript{2}
partial pressure inside the chamber was maintained at $5\times10^{-3}$
mbar and the substrate temperature was 500 \textsuperscript{o}C for
all the samples. The number of laser pulses were varied to grow thin
films of different thicknesses. A 4-circle high resolution XRD (HRXRD)
(Rigaku Smartlab, Japan) was employed to investigate the structural
quality and also to determine the thickness of the samples from X-ray
reflectivity (XRR). These experiments were performed using a Ge (220)
monochromator and the Cu K\textsubscript{$\alpha$} source. Temperature
dependent electrical and magnetotransport measurements were carried out using
a physical property measurement system (PPMS, Quantum Design) in the
temperature range of 4 \textendash{} 300 K.

\section{Experimental results}

\subsection{X-ray reflectivity}

\begin{figure}
\includegraphics[width=8.5cm]{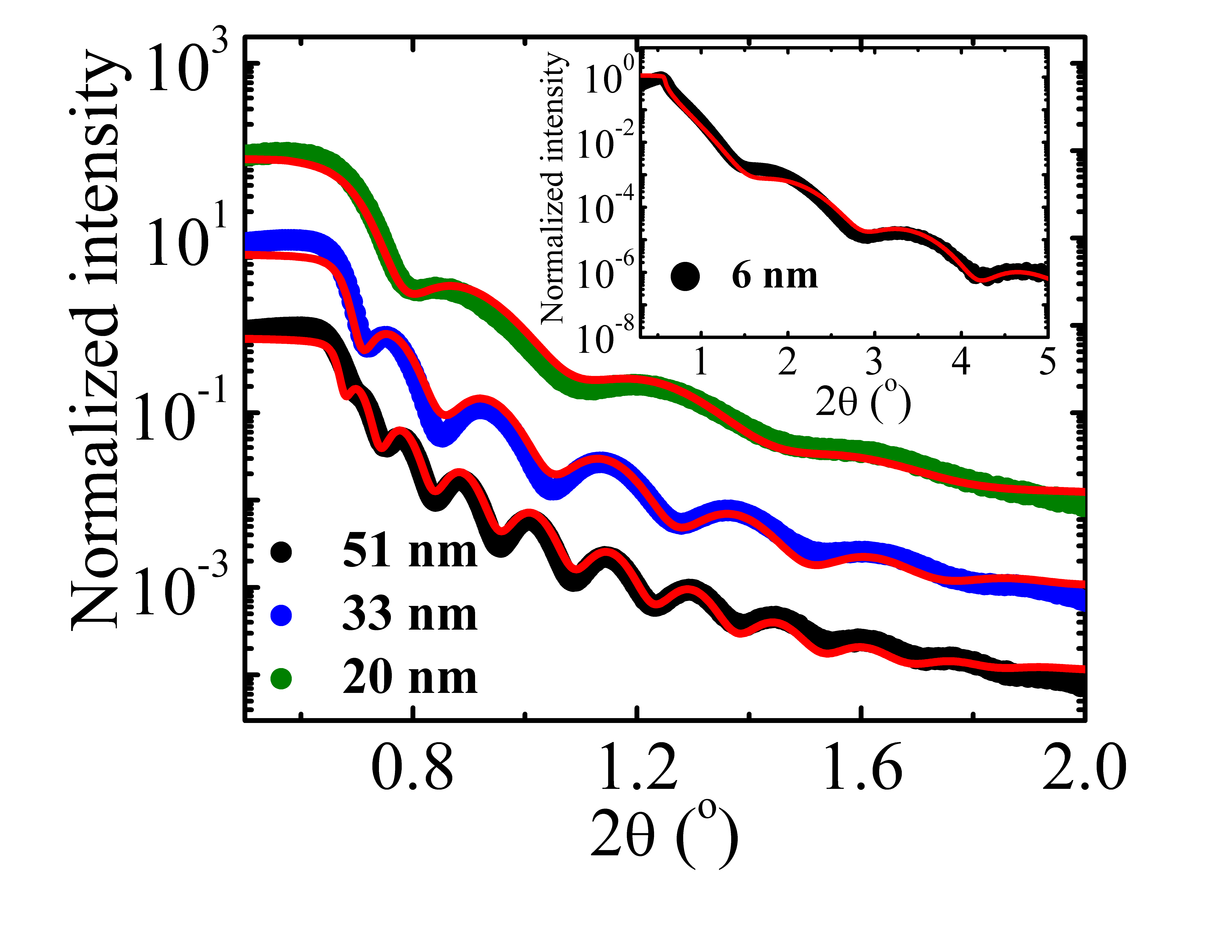}

\caption{X-ray reflectivity profiles of Ga:ZnO thin films grown on sapphire
using PLD. Inset shows XRR profile of 6 nm film in larger x-axis range. The clear oscillations indicate that the film surface and
the interface are very smooth. The rms roughness of the samples is
below 0.7 nm. Red lines are fitting of experimental data using globalfit
software. Each profile is multiplied by a factor of 10 for easy visualization.\label{fig:XRR}}
\end{figure}

To precisely determine the film thickness, we have carried out X-ray reflectivity (XRR) on Ga:ZnO films. Figure \ref{fig:XRR} displays XRR profile of Ga:ZnO thin films
of different thickness grown on c-sapphire substrate. XRR data was
fitted using Globalfit software (Rigaku) to estimate the thickness
and roughness of the films. It can be observed from the Fig. \ref{fig:XRR} that the fitted data are in excellent agreement with the experiment.
The thickness values as obtained from the fit are indicated in the
figure and have been used to analyze the electrical transport data.
The root mean square (rms) roughness of the surface of each film lies
below 0.7 nm implying an extremely smooth nature of Ga:ZnO films.

\subsection{X-ray diffraction}

\begin{figure}
\includegraphics[width=8.5cm]{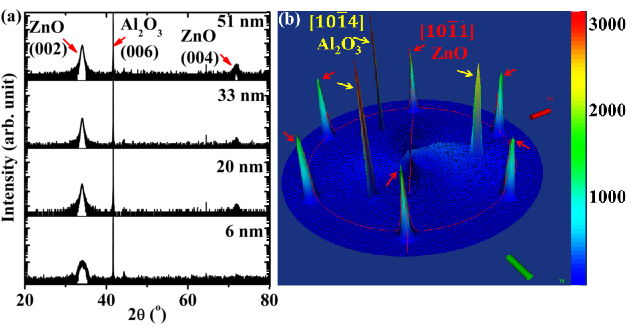}
\caption{XRD patterns of Ga doped ZnO thin films grown on saphhire substrates
by PLD. The sharp peak at 41.7\textsuperscript{o} corresponds to
Sapphire (006). (b) 3D pole figure of Ga:ZnO thin film of thickness
20 nm. Peaks corresponding to ZnO and sapphire are marked with red and yellow arrow. It can be seen that substrate and film peaks are 30\textsuperscript{o}
apart.\label{fig:XRD}}
\end{figure}

It is essential to check the crystallinity and the epitaxial nature
of Ga:ZnO films particularly in low thickness. Figure \ref{fig:XRD}(a) depicts the
$2\theta/\omega$ scan of the films with different thickness grown
by PLD on c-sapphire substrate. Only (002) and (004) peaks of ZnO
were observed along with the substrate (006) peak.
This confirms that films are highly textured along the c-axis. The absence
of additional peak confirms that there are no Ga\textsubscript{2}O\textsubscript{3}
or other phases present in the samples. As film thickness decreases, full width at half maxima (FWHM) of (002) peak increases and the intensity of the peak decreases monotonically. The FWHM of 51 and 6 nm film is 0.56\textsuperscript{o} and 1.6\textsuperscript{o} respectively\cite{Dong2007GaZnO_XRD}. The broadening of (002) peak in the thinnest film can be attributed to the lattice disorder and strain present in the film. Moreover, to confirm the epitaxial nature of the
films, pole figure measurements were carried out on all the films. Pole figure measurement was performed on {[}10-11{]} set of planes of ZnO
and Fig. \ref{fig:XRD}(b) displays the 3D plot of pole figure of 20 nm thick sample. Six equally spaced peaks were obtained
which confirm the six fold symmetry of ZnO\cite{Opel2014_ZnO_Epitaxy, Snure2007_MIT_Ga_conc_Mott_limit}. In our films, we do not observe any 30\textsuperscript{o} in-plane domain rotation of ZnO on c-sapphire substrate\cite{Opel2014_ZnO_Epitaxy}.

\subsection{Resistivity}

\begin{figure}
\includegraphics[width=8.5cm]{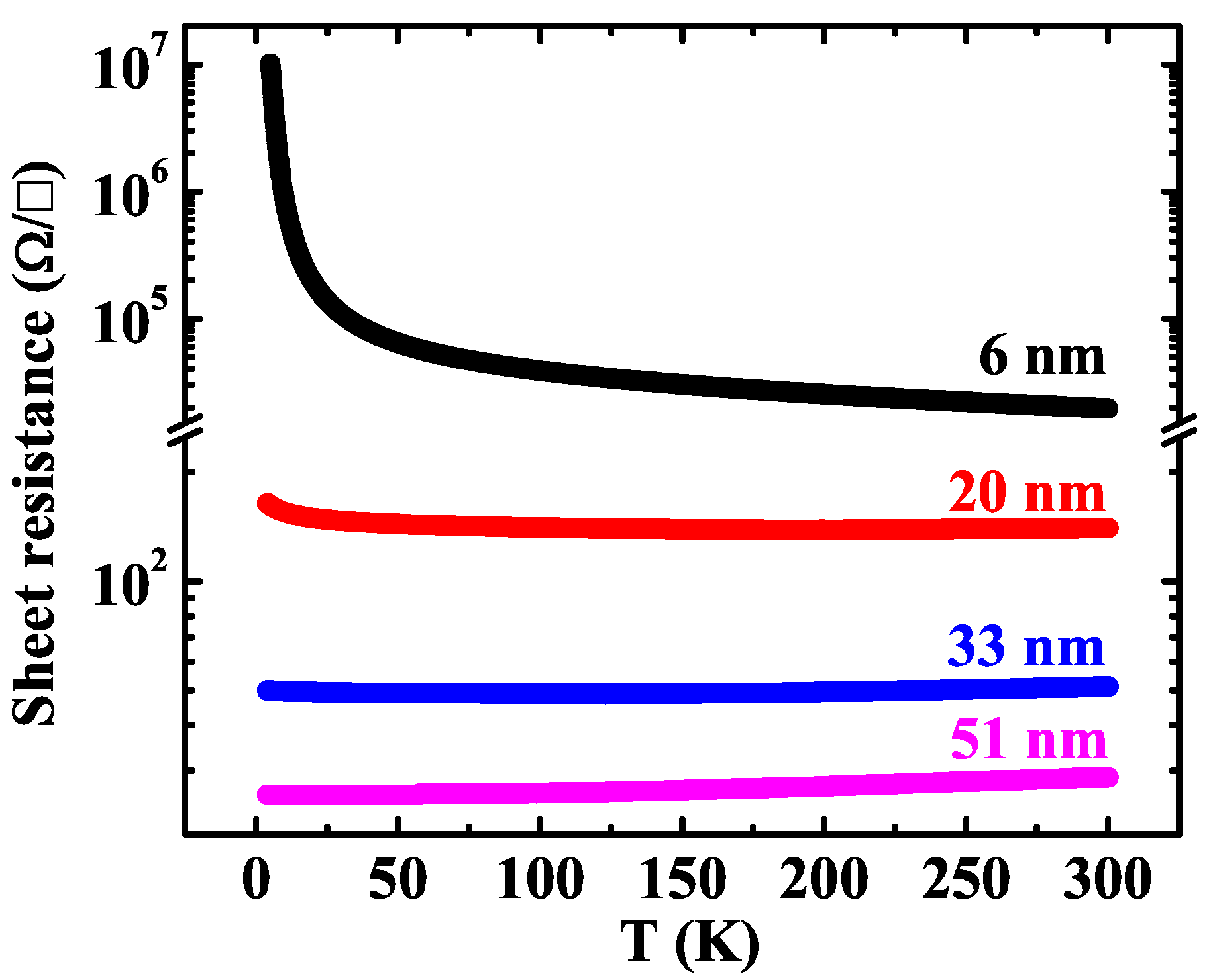}
\caption{Temperature dependent sheet resistance of Ga doped ZnO thin films
of different thickness. Note that the y- axis is in
logarithimic scale. As thickness decreases, the sheet resistance increases
monotonically. The film of thickness 6 nm shows an insulating behavior
whereas others exhibit metallic type of conduction.\label{fig:sheet resistance}}
\end{figure}

To investigate the thickness driven metal-insulator transition (MIT)
in Ga:ZnO films, temperature (T) dependent sheet resistance measurements
on all the samples were carried out using standard 4-probe method. Irrespective of T, the sheet
resistance ($R$) increases as film thickness is reduced.
The values of $R$ for 51, 33 and 20 nm thick films are 29 , 51 and 140 $\Omega$ respectively. With further decrease in film thickness, there
is a sharp increase in $R$ as shown in Fig. \ref{fig:sheet resistance}. At 300 K, $R$ of 6 nm film is found to be 19.7 k$\Omega$ . Similar behavior is observed in Si doped ZnO thin films by Das et al\cite{Das2014_SiZnO_MIT}. Sheet resistance (R), resistivity ($\rho$) and carrier concentration ($n$) of Ga:ZnO films measured at 300 K are listed in the Table \ref{tab:electrical paramaters}.
\begin{table*}
\caption{The list of sheet resistance ($R_{300}$, measured at 300 K), resistivity ($\rho_{300}$),  $\rho_{4}/\rho_{300}$ and electron concentration ($n$) values of Ga:ZnO thin films of different thickness.
\label{tab:electrical paramaters}}
\begin{tabular}{c@{\hskip 1cm}c@{\hskip 1cm}c@{\hskip 1cm}c@{\hskip 1cm}c}
\hline 
\hline
Film thickness (nm) & $R_{300}$($\Omega$) &  $\rho_{300}$($\Omega$-cm)  & $\rho_{4}/\rho_{300}$  & $n$(cm\textsuperscript{-3}) \tabularnewline
\hline 
51 & 29 & $6.6\times10^{-4}$ & 0.92 & $1.5\times10^{21}$ \tabularnewline
33 & 51 & $7.7\times10^{-4}$ & 0.97 & $1.3\times10^{21}$ \tabularnewline
20 & 140 & $1.3\times10^{-3}$ & 1.2 & $8.6\times10^{20}$ \tabularnewline
6 & 19667 & 0.05 & 188 & $7.1\times10^{20}$ \tabularnewline
\hline 
\hline
\end{tabular}
\end{table*}
The carrier concentration is found to be $1.3\times10^{21}$ and $7.1\times10^{20}$ cm\textsuperscript{-3} for 51 and 6 nm films respectively. Figure \ref{fig:sheet resistance} depicts the behavior
of sheet resistance ($R$) as a function of temperature for all Ga:ZnO
films of different thickness. We find that there is a weak dependence
of $R$ on temperature for thicker films (20 nm and above). However, in the case of 6 nm thick film,
there is a sharp increase in $R$ below 50 K which is the characteristic feature of an insulator.

\begin{figure}
\includegraphics[width=7cm]{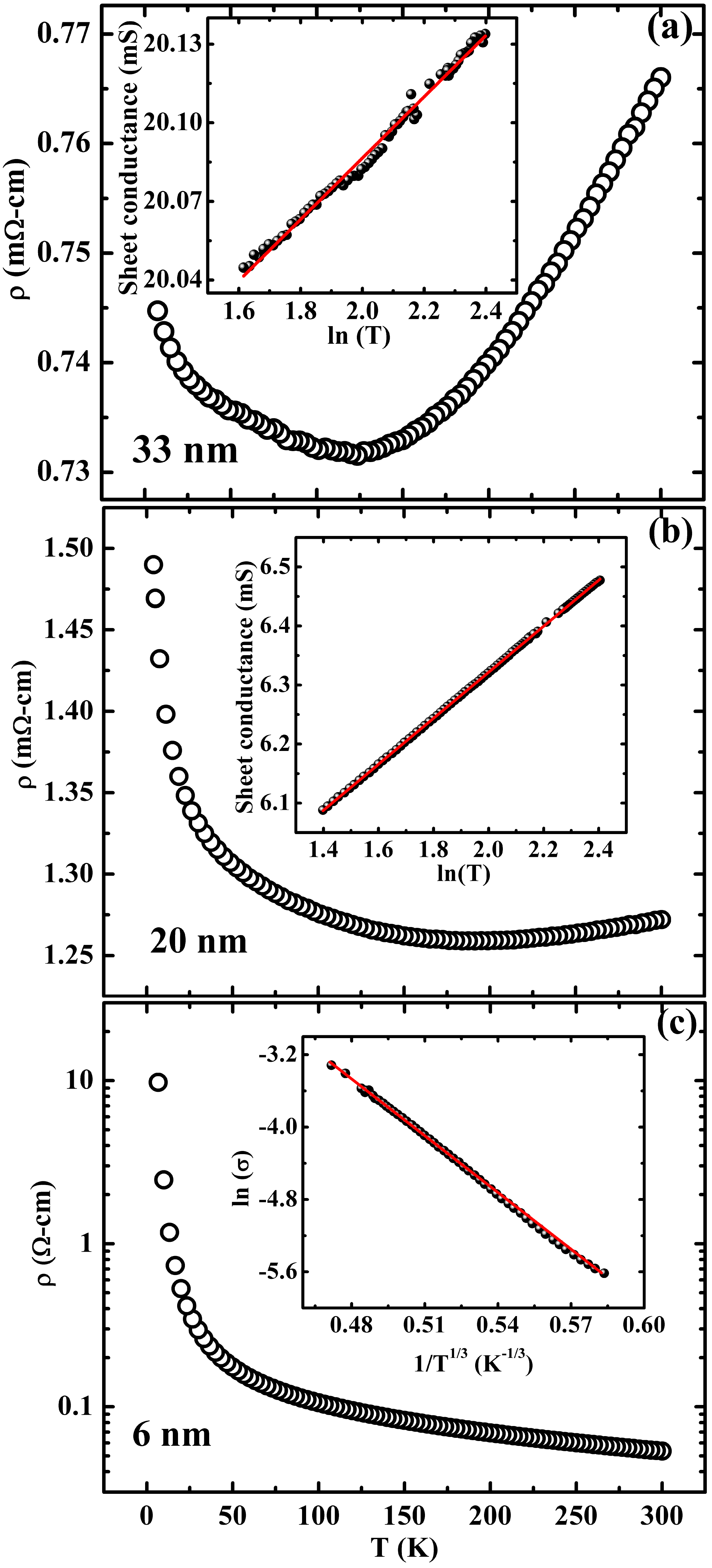}
\caption{Temperature dependent resistivity plot of (a) 33, (b) 20 and (c) 6 nm thick Ga:ZnO films. Insets of (a) and (b) represent $\sigma$ \textit{vs.} $ln(T)$ plot of 33 and 20 nm thick films respectively. Inset of (c) shows the linear behavior
of $ln\left(\sigma\right)$ \textit{vs.} $1/T^{1/3}$ at low temperature regime below 50 K which indicates the existence of 2D Mott\textquoteright{}s VRH mechanism. Red lines are linear fitting to the experimental data.\label{fig:resistivity}}
\end{figure}

For further analysis of electrical transport properties, we have plotted resistivity
($\rho$) as a function of temperature as shown in Fig. \ref{fig:resistivity}. At
300 K, $\rho$ of the 51 nm thick film is found to be 0.66 m$\Omega$-cm which is very close to the earlier reports\cite{Bhosle2006_MIT_Ga_conc}. For 33 and 20 nm films, $\rho$ at 300 K is 0.77 and 1.3 m$\Omega$-cm respectively. Figures \ref{fig:resistivity}(a) and (b) display the plot of $\rho$ as a function of temperature for 33 and 20 nm films. As temperature
is lowered down from 300 K, $\rho$ decreases and reaches to a minimum
value. On further reduction in temperature below the transition temperature (T\textsubscript{MIT}), $\rho$ increases. The resistivity minimum
is generally observed in degenerate semiconductors\cite{Das2014_SiZnO_MIT}. Both the samples show positive temperature coefficient of resistivity (TCR) at high temperature
regime and below the T\textsubscript{MIT}, TCR becomes negative. Transition temperature (T\textsubscript{MIT}) is observed
at 125 K and 190 K for the films with thickness 33 and 20 nm respectively.
It should be noted that T\textsubscript{MIT} increases as the film thickness is reduced.
The negative TCR below the T\textsubscript{MIT} can be explained with the
frame of weak localization (WL) phenomena. In the case of weak
localization, due to quantum interference, the electronic wave functions
diffuse and thereby enhance the backscattering which in turn increases
the resistivity. 

The conductivity in the weak localization regime
for a 2D system is given as \cite{Lee1985_RMP_disorder, Scherwitzl2011_LNO_MIT_2D_WL},
\begin{equation}
\sigma=\sigma_{0}+\frac{pe^{2}}{\pi h}ln\left(T/T_{0}^{'}\right)
\end{equation}

where, $\sigma_{0}$ is Drude conductance and $p$ is a parameter that depends on the nature of inelastic dephasing
mechanism. The value of $p$ is one for electron-electron collision
and 3 for electron-phonon collision\cite{Scherwitzl2011_LNO_MIT_2D_WL}. The constant $T_{0}^{'}$
is related to the mean free path for elastic collision. Insets of Fig. \ref{fig:resistivity} (a)
and (b) displays the $\sigma$ \textit{vs.} $ln(T)$ plot of the 33 nm
and 20 nm films at low temperature regime ($<$T\textsubscript{MIT}) respectively.
The linear behavior of $\sigma$ with $ln(T)$ can be observed
from these insets. The slopes are found to be $3.8\times10^{-4}$
and $5.2\times10^{-4}$ for the 33 nm and 20 nm samples respectively. Therefore, we infer that the transport follows the 2D weak localization behavior
in these samples. The resistivity value and the positive TCR suggests that the films with thickness
20 nm and above exhibits metallic type conduction. Moreover, the resistivity
at 5 K and 300 K are very close in these samples which also indicate
a metallic type behavior\cite{Mukherjee2017_IGZO}.

The temperature dependent resistivity plot of 6 nm thick film is reverse
to the trend shown by thicker samples.
From Fig. \ref{fig:resistivity}(c) we observe negative temperature coefficient of resistivity
(TCR) throughout the temperature range which conclusively predicts
the insulating behavior. The slope at low temperature regime ($<50$
K) is higher than the high temperature regime as can be realized from
Fig. \ref{fig:resistivity}(c). The two different slopes can be attributed to the two
different conduction mechanisms present in the two different temperature
scales. Generally at high temperature regime, the transport is mainly
governed by thermally activated band conduction. For thermally activated
band conduction, the conductivity is given by $\sigma\left(T\right)=\sigma'exp\left(-E_{a}/k_{B}T\right)$
where, $\sigma'$ is the pre-exponent factor, $E_{a}$ is the activation
energy. The Arrhenius fitting of the conductivity data yields the activation
energy as 27 meV. Resistivity sharply increases at the low temperature
regime below 50 K. At low temperature, the thermal energy is insufficient
to activate the carriers. Therefore, the conduction process mainly takes place
through the variable range hopping mechanism. In this process electrons
can hop from one site to other sites that are close to the Fermi level
irrespective of their spatial separation. The relation between electrical
conductivity and temperature for VRH mechanism in 2D system is given
by\cite{Stepina2009_2D_VRH},
\begin{equation}
\sigma\left(T\right)=\sigma'_{0}exp\left(-\frac{T_{0}}{T}\right)^{1/3}
\end{equation}

where, $\sigma'_{0}$ is a pre-exponent constant and $T_{0}$ is the
characteristic temperature. Inset of Fig. \ref{fig:resistivity}(c) displays the variation
of $ln\left(\sigma\right)$ \textit{vs.} $1/T^{1/3}$ of 6 nm thick sample
in the temperature range of 50 to 10 K. The linear fitting (red solid
line in Fig. \ref{fig:resistivity}(c)) of the plot indicates that the VRH mechanism is
dominant in the conduction below 50 K.
\subsection{Magnetoresistance}
\begin{figure}
\begin{centering}
\includegraphics[width=8cm]{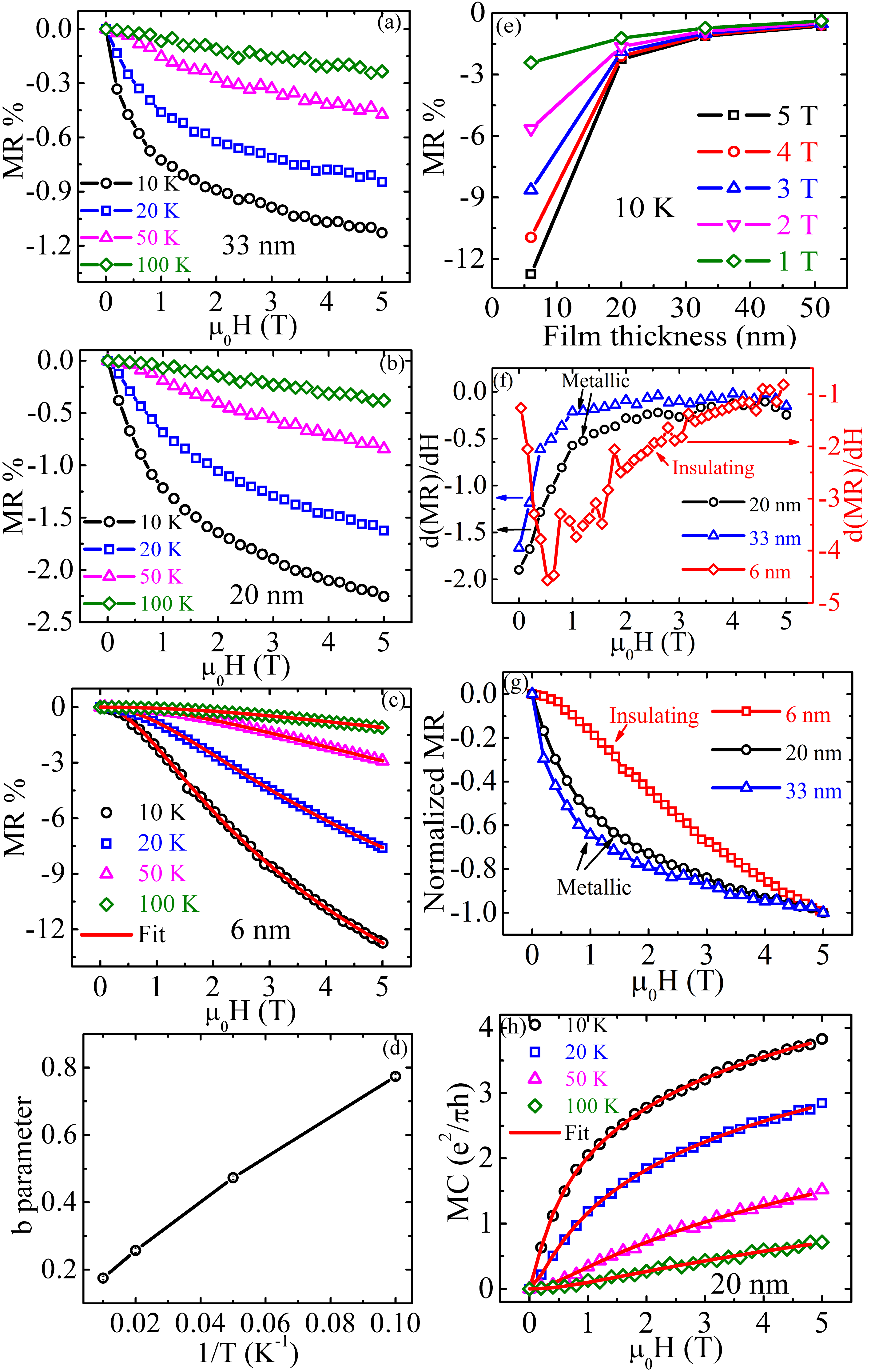}
\par\end{centering}

\caption{Magnetoresistance of (a) 33, (b) 20 and (c) 6 nm thick Ga:ZnO
film. Solid red lines in (c)
are fit of the experimental data according to the Eq. (\ref{eq:MR_Khosla-ve}). (d) Fitting parameter b vs. 1/T plot. The parameter b scales inversely with temperature as expected from Eq. (\ref{eq:b-parameter}). (e) MR\% as a function of film thickness at 10 K. As thickness decreases, MR amplitude increases in all temperatures.
Films with thickness 20 nm and above (metallic) show a small MR \%
compared to the 6 nm (insulating) film.  (f)The
first derivative of MR\% with respect to the applied field and (g) normalized MR of 33, 20 and 6 nm thick films. For metallic samples (20
nm and above), amplitude of first derivative is more at low field
and decreases as field is increased. This indicates that at low field
in these samples, the variation of MR\% is higher compared to the
high field regime. However, for 6 nm thick sample, amplitude of first
derivative increases with field and after reaching a maximum value
it decreases again.(h)Magnetoconductance behavior of 20 nm thick film which agrees well with 2D weak localization model. Red line is the fit according to the Eq. (\ref{eq:2D-WL})\label{fig:MR}}
\end{figure}
The resistivity data discussed in the last section not only infers metal insulator transition (MIT), but also it provides indication of strength of electron localization. The resistivity data of 6 nm film fits well with the proposed VRH model suggesting strong localization (SL). At the same time, the small upturn at low temperature regime for thicker (20 nm and above) films hints toward weak localization (WL) which agrees well with 2D WL model. In order to substantiate the MIT and to verify the nature of localization inferred from the resistivity data, we have carried out magnetoresistance measurement and the results are displayed in Fig. \ref{fig:MR}. As expected, from Fig. \ref{fig:MR} (a-c), we find that irrespective of the film thickness, the MR became more negative with decrease in temperature. The negative MR has been observed on Ga doped ZnO thin films by Reuss \textit{et al}\cite{Reuss2005_Gadoped_MR}. While the MR vs. H is qualitatively similar, there is a large quantitative difference between metallic and insulating samples. As film thickness decreases, magnitude of MR increases which can be realized from Fig. \ref{fig:MR}(e). At 10 K with 5 T field, the maximum MR observed in 20 and 33 nm thick samples is 1.3 and 2.4 \% respectively. Generally, in WL regime, a small magnitude ($\thicksim$1-2 \%) of MR is observed due to the suppression of WL in the presence of magnetic field\cite{Andrearczyk2006_NgMR_WL, Mukherjee2017_IGZO}. 

To understand the variation of MR on applied magnetic field, we have plotted (d(MR)/dH) and normalized MR in Figs. \ref{fig:MR}(f) and (g) respectively. From the normalized plot, it can be observed that for the metallic samples, the magnitude of MR increases sharply at low field regime ($<0.5$ T). Therefore, we observe a large value of d(MR)/dH at small H (see Fig. \ref{fig:MR}(f)). On further increase in field strength, d(MR)/dH approaches to saturation indicating smaller and constant change in MR with field. This is because a small magnetic field can suppress the weak localization effect\cite{Mukherjee2017_IGZO}. Further increasing the field strength yields a lower change in MR. Therefore, the change in MR with respect to the field is more prominent at low magnetic field compared to the higher field.
To fit the experimental data, we have applied 2D WL model for 20 nm and above thick films. We have calculated the magnetoconductance (MC) for the metallic
samples defined as $MC=\sigma\left(H\right)-\sigma\left(H=0\right)$.
In these samples, temperature dependent resistivity measurement confirmed
the presence of weak localization at low temperature regime. In 2D
system, MC due to weak localization effect is given by \cite{Lee1985_RMP_disorder, Scherwitzl2011_LNO_MIT_2D_WL},
\begin{equation}
MC=\frac{e^{2}}{\pi h}\left[\varphi\left(\frac{1}{2}+\frac{1}{x}\right)+ln\left(x\right)\right]\label{eq:2D-WL}
\end{equation}
where, $e$ is electronic charge, $h$ is the Planck's constant, $\varphi$
is Digamma function. The term $x$ is defined as $x=l_{in}^{2}\frac{4eH}{h}$
where, $l_{in}$ is the inelastic scattering length. The magnetoconductance
data of Ga:ZnO films with thickness 20 nm and above are fitted to
the Eq. (\ref{eq:2D-WL}). Figure \ref{fig:MR}(h) shows magnetoconductance behavior of 20 nm thick film. It can be observed that the experimental data is well agreed with the 2D WL model. The inelastic scattering length was found to be 65 nm at 10 K and 15 nm at 100 K in this film.

Unlike to the metallic samples, the MR of 6 nm thick sample is different in two aspects. First the magnitude is much larger than that of metallic samples and the dependence of MR on H is also different. A maximum value of 12.7 \% MR is observed in 6 nm thick sample which shows strong localization behavior in the resistivity measurement. This magnitude of MR agrees with the MR observed in other doped ZnO films exhibiting strong insulating behavior\cite{Agrawal2014_Mgdoped_MR, Mukherjee2017_SbZnO}. For insulating sample d(MR)/dH vs. H curve exhibits a minimum at a magnetic field of 0.5 T at 10 K. Similar behavior is observed in La\textsubscript{0.7}Ca\textsubscript{0.3}MnO\textsubscript{3} system in the insulating regime\cite{Pekala2008_LCMO}. Below this minimum ($<0.5$ T), magnitude of d(MR)/dH increases with field. After reaching the maximum, the magnitude of d(MR)/dH decreases which indicates a slower rate of MR with field. In strong localization limit negative MR is explained by the scattering center. Upon application of a magnetic field, scattering centers align among themselves. As, a high field is required to align the scattering centers, d(MR)/dH first increases with field and after reaching the maximum value it starts decreasing. It has been observed by many\cite{Mukherjee2017_SbZnO} that in the strong localization regime, negative MR follows the semi-emperical Khosla-Fischer model\cite{Khosla1970_MR}. We have analyzed the MR data of 6 nm thick film with a semi-empirical model proposed by Khosla and Fischer. According to this model, negative MR is given by\cite{Khosla1970_MR},
\begin{equation}
MR=\frac{\triangle\rho}{\rho}=-a^{2}ln\left(1+b^{2}B^{2}\right)\label{eq:MR_Khosla-ve}
\end{equation}

The parameters a and b are defined as

\begin{equation}
a^{2}=A_{1}JN\left(E_{F}\right)\left[S\left(S+1\right)+\left\langle M^{2}\right\rangle \right]
\end{equation}

and

\begin{equation}
b^{2}=\left[1+4s^{2}\pi^{2}\left(\frac{2JN\left(E_{F}\right)}{g}\right)^{4}\right]\left(\frac{g\mu_{B}}{\alpha k_{B}T}\right)^{2}\label{eq:b-parameter}
\end{equation}

where constant $A_{1}$ is a measure of spin based scattering, $J$
is the exchange integral, $N(E_{F})$ is the density of states at
the Fermi energy, $S$ is the spin of the localized moment, $M$ is
the average magnetization, $g$ is the g-factor and $\alpha$ is a
numerical constant. The MR data of 6 nm thick Ga:ZnO film was fitted
using the Eq. (\ref{eq:MR_Khosla-ve}). Figure \ref{fig:MR}(c) suggests that experimental data are in good agreement with the Khosla
and Fischer model. The fitting parameter $b$ is plotted in Fig. \ref{fig:MR}(d) and it varies inversely with temperature as expected from Eq. (\ref{eq:b-parameter})\cite{Gacic2007_twobandmodel}. The other parameter $a$ slightly decreases with temperature\cite{Mukherjee2017_SbZnO, May2004_khosla_b, Peters2010_khosla_b}.

\section{Lattice disorder and electronic structure}

\begin{figure}
\begin{center}
\includegraphics[width=8cm]{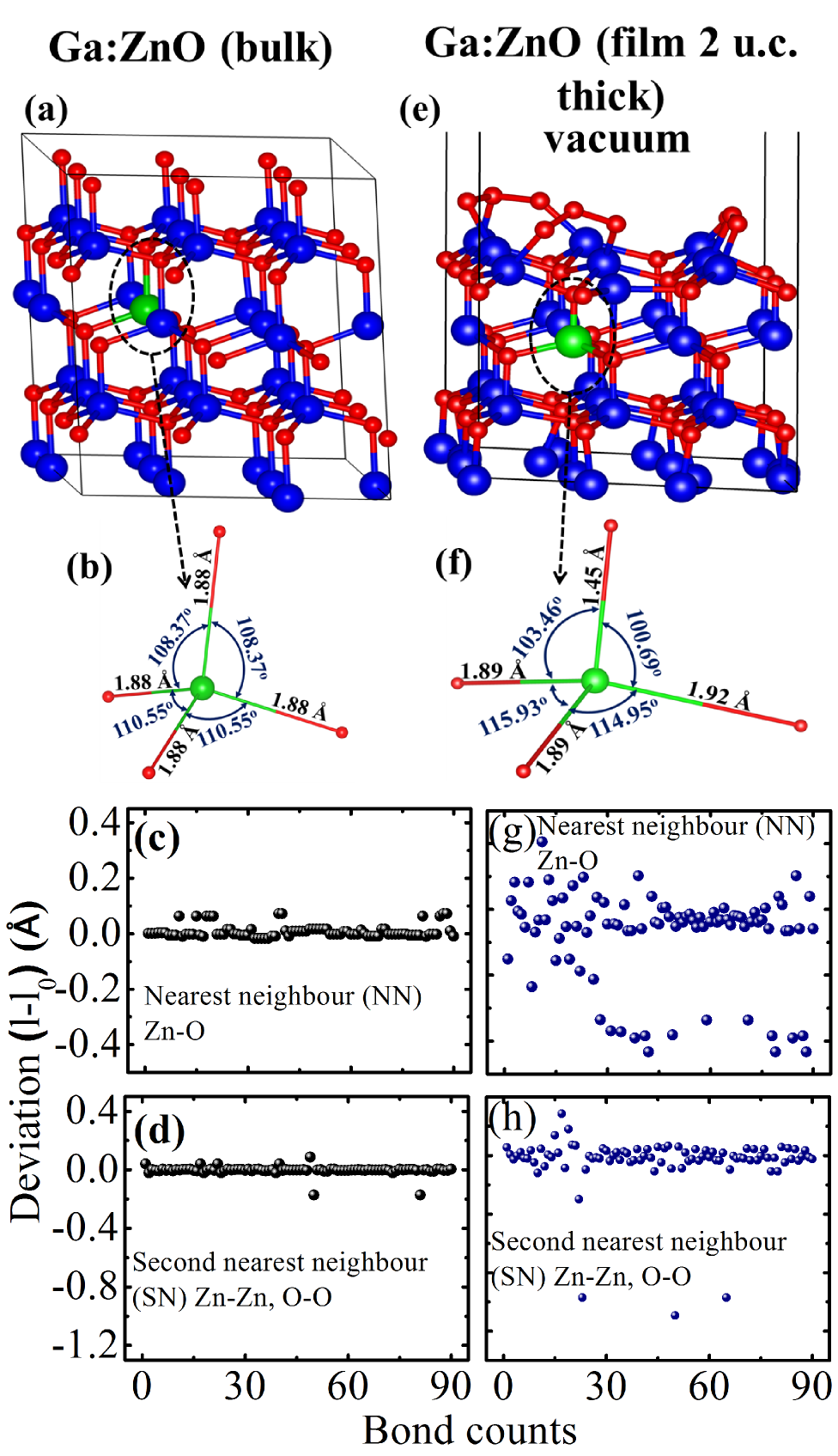}
\end{center}
\caption{Strength of crystallinity in Ga doped ZnO in bulk (a, b, c and d)
and in two unit cell thick (1nm) film (e, f, g and h). While (a) and
(e) show the fully relaxed structure, (b) and (f) indicate the distortion
occurring at the immediate neighborhoods of the dopant. Figs. (c)
and (g) plot the deviation of all nearest neighbor (NN) bond lengths
from that of the undoped bulk compound ($l-l_{0}$) . More
the deviation, both in count and magnitude, stronger the disorder.
Similarly, Figs. (d) and (h) plot the deviation of second neighbor
(SN) bond lengths. Total number of NN and SN bonds for the considered
$3\times3\times2$ supercell are 90.\label{fig:Structural-disorder}}
\end{figure}

To examine the electronic contribution to the transport properties
of Ga:ZnO film, in this section we analyzed the band structure  of this doped system obtained from the DFT calculations.   
While experimentally synthesized samples have 1$\%$ Ga doping, due to computational limitations, theoretical studies are performed with 3$\%$ Ga doping. The doping is realized by replacing one Zn atom by Ga in a $3\times3\times2$ supercell as shown in Fig. \ref{fig:Structural-disorder}(a).  Also due to computational limitations, calculations are performed only on bulk Ga:ZnO and two unit cell thick (1 nm) film along (001) (see Fig. \ref{fig:Structural-disorder}(e)). While the former can represent the experimentally synthesized thicker films, the latter closely stands for the  6 nm thin film. The film was constructed by introducing 15 \AA vacuum on the surface of 2 unit cell thick supercell as shown in the Fig. \ref{fig:Structural-disorder}(e). The ground state structure is realized through complete relaxation of the experimental structure of bulk ZnO\cite{Sans2009_doped_ZnO_BS}. 

The calculations are carried out using the Vanderbilt
ultra-soft pseudopotential\cite{Vanderbilt1990_Pseudopotential} and plane wave basis sets as implemented in Quantum Espresso\cite{Giannozzi2009_QE} and within the frame work of PBE-GGA exchange correlation approximation\cite{Perdew1996_PBE}. The kinetic energy cutoff to fix the number of plane waves is taken
as 30 Ry. The Brillouin zone integration is performed using the Gaussian method on a $16\times16\times16$ k-grid for bulk and $16\times16\times1$ k-grid for film.

With introduction of impurities (vacancy, dopant etc.), crystalline
disorder occurs and the extent of disorderness primarily depends on
the charge state and ionic radii of the impurities. Subsequently,
the disorder introduces new localized states and thereby alters the
transport properties of the compound.
To explain this aspect, we have measured the crystalline disorder
of Ga:ZnO both in bulk and thin film. The results are shown in Fig. \ref{fig:Structural-disorder}. Figures \ref{fig:Structural-disorder}(a) and (e) respectively show the fully relaxed
structure of bulk and film and it is visible that in film disorder is more than that in the bulk. Moreover, from Fig. \ref{fig:Structural-disorder}(b) we find that for bulk Ga:ZnO, the impurity Ga-O bond in the nearest neigborhood is isotropic with length 1.88 \AA. However, in the case of film (see Fig. 6(f)), the Ga neighborhood is highly anisotropic with Ga-O bond length varies from 1.45 to 1.92 \AA.
 
 To provide a quantitative measure of disorderness in the doped
system, we estimated the deviation of nearest neighbor (NN) and second
neighbor (SN) bond length from that of the undoped bulk ZnO. The results
are presented in Figs. \ref{fig:Structural-disorder}(c-d) and \ref{fig:Structural-disorder}(g-h). We find that out of 64 possible
NN bonds, a maximum of 7 bonds are deviated by 0.1 \AA from the
original value in the case of bulk compound. However, in the case
of film nearly 56 bonds are deviated significantly and the deviation
lies in the range of -0.5 to 0.2 \AA. The same can be realized from
the SN bond lengths even though the number of deviations is comparatively
less. Therefore, collectively we find that the extent of disorder
in film is significant and likely to impact the transport properties which can be understood from the band structure and density
of states. 

\begin{figure}
\includegraphics[width=7cm]{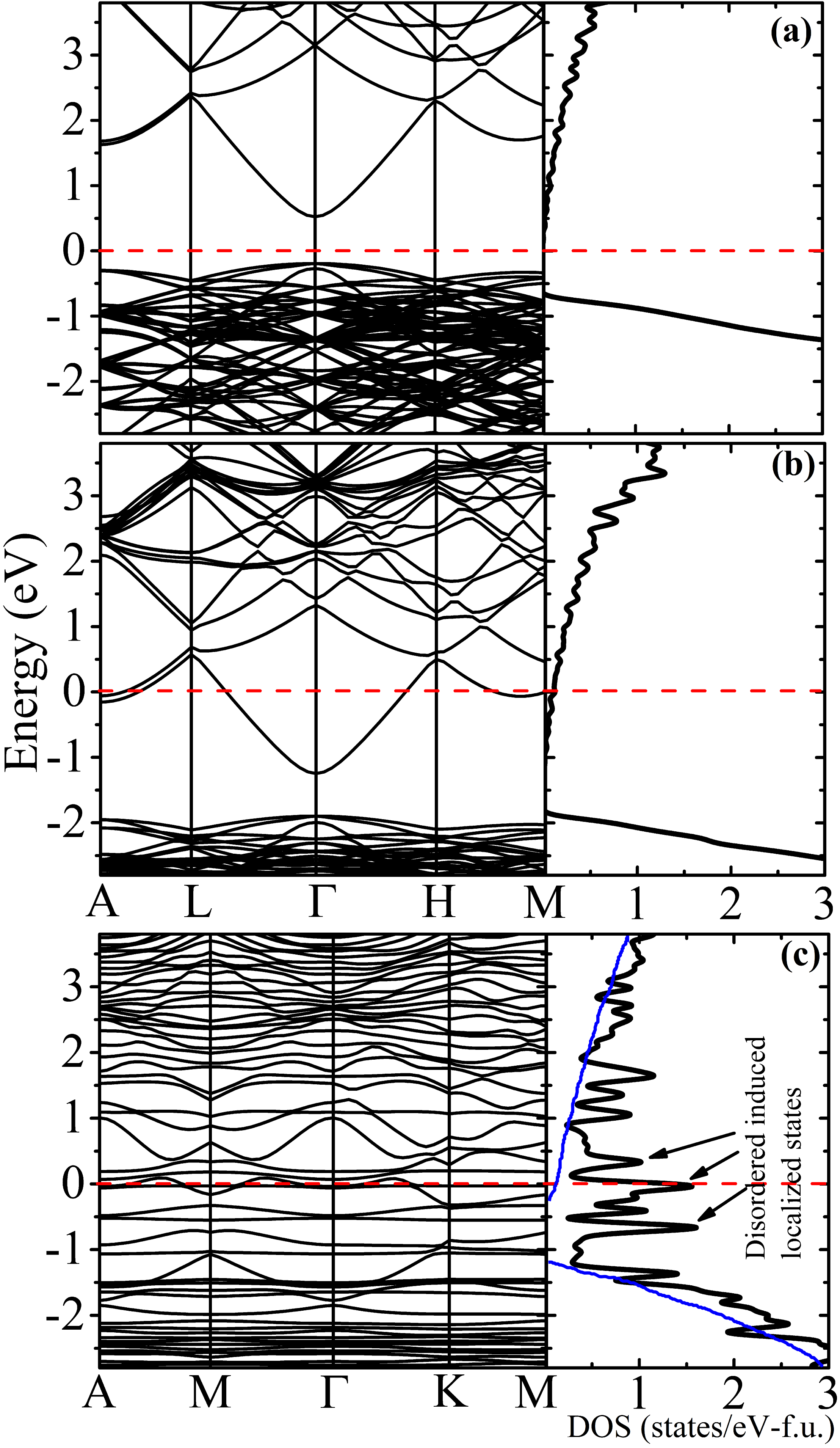}\caption{ Band structure and DOS of (a) bulk ZnO, (b) bulk Ga:ZnO and (c) 2 unitcell thick Ga:ZnO film. The band structure for the bulk compounds are plotted using the Brillouin zone of a 3$\times$3$\times$2 supercell.For the film, the BZ is that of a 3$\times$3$\times$(2+vacuum) supercell (see Fig. \ref{fig:Structural-disorder}(e)).
As expected the bulk doped band structure represents  n-type degenerate semiconductor with a dispersive parabolic band crossing the Fermi level (E$_F$). In the case of film, the strong disorder introduces highly localized states, as indicated, to decrease the conductivity. The blue curve in the bottom is a guide to eye to distinguish between the delocalized and localized DOS of Ga:ZnO in bulk and film respectively. The high symmetry points defining the Brillouin zone of the bulk are A (0,0,0.47), L (0, 0.5, 0.47),$\Gamma$ (0, 0, 0), H (0.33, 0.33, 0.47),  and M (0, 0.5, 0). For 2 unit cell thin film, the considered high symmetry points are A (0,0,0.5), M (0, 0.5, 0), $\Gamma$ (0,0,0), and K (0.33, 0.33, 0.47).\label{fig:Band-structure-DOS}}
\end{figure}

ZnO is a known semiconductor with the Fermi level lies between the occupied O-p dominated valence bands and Zn-4s dominated dispersive conduction band.  Figure \ref{fig:Band-structure-DOS} (a) shows the band structure of pure ZnO plotted using the Brillouin zone of a 3$\times$3$\times$2 supercell. The lower conduction band is highly dispersive and with a large band width of 7 eV. Due to band folding this band consists of several subbands which are degenerate along A-L and also at H.  With Ga doping, both the orbital and crystal symmetry break along with a minor crystal disorder which is discussed in Fig. \ref{fig:Structural-disorder}. Such disorder and symmetry breaking lead to the lifting of the subband degeneracy (See the band dispersion in the range 1-3 eV for pure and -0.5 to 1 eV for doped sytem with respect to E$_F$) which can be seen from Fig. \ref{fig:Band-structure-DOS}(b). Now the Ga-p dominated lower subband, which is parabolic with width 2 eV and separated from the upper subbands of the conduction spectrum, is partially occupied to make the system a n-type degenerate semiconductor. The parabolic behavior of the impurity band is in good agreement with our experimental study where
we have shown that the thickest film (51 nm) exhibits positive temperature coefficient of resistivity like a metal. The density of states in Fig \ref{fig:Band-structure-DOS}(b) reflects the same where the tail
of the conduction band is smooth and well below E$_F$.

As discussed earlier, the crystalline disorder in film is very profound in the film. As a consquence, the bulk covalent band gap disappears. Instead several localized state appears below and above E$_F$ as can be seen from the bands and DOS of Fig \ref{fig:Band-structure-DOS}(c).
 Since, one electron is donated by Ga, it is natural to see a partially occupied
band. Howevere, unlike to the bulk case, such a band is  of narrow band width due to localization effect.
From Fig. \ref{fig:Band-structure-DOS}(c) we observe that the
band width of this partially occupied band is approximately 0.2 eV which is
almost one order less than that of the doped bulk system signifying the strength
of the crystal disorder.

In earlier studies on hydrogenated amorphous Si\cite{Mott1980_gap_states}\cite{Okamoto1977_gap_states}, such
localized states in the vicinity of the E\textsubscript{F} have been
found. These states are  reported to be responsible for the insulating behavior, mediated through the VRH mechanism, of these compounds. The localized states of Ga:ZnO film
here resembles to that of amorphous Si and favor the insulating behavior. The experimentally observed insulating behavior of 6 nm film resonates with the lattice disorder led insulating phenomena of 2 unit cell thick film. We may note that Fig. \ref{fig:XRD} indicates a broad peak in the XRD pattern of the 6 nm film as opposed to the narrow peaks shown by the 20 nm and above thicker films.

\section{Summary and conclusions}

To sumarize, structural, electrical and magnetotransport properties
of PLD grown Ga doped ZnO thin films were extensively studied and
analyzed. Metal-insulator transition was observed as a function of
the film thickness. While the sample with thickness of 20 nm and above exhibits
metallic type of conduction, the film of 6 nm thick is
found to be insulating. Temperature dependent resistivity and magnetoresistance measurements further confirm the presence of weak localizatin in 20 nm and above thicker samples. However, strong localization is observed in 6 nm thick film as resistivity data follow Mott-VRH model. Moreover, the magnetoresistance data is well matched with Khosla-Fischer model to further substantiate the strong localization behavior in 6 nm film. Our structural relaxation calculations, carried out using density functional theory, suggests that the dopant as well surface reconstruction introduce larger disorder in thinner samples. As a consequence, the system has disorder induced localized states in the vicinity of Fermi level. The donor electrons occupy some of these localized states to weaken the mobility and exhibit insulating behavior.
\begin{acknowledgments}
Authors thank Department of Science and Technology (DST),
New Delhi, that facilitated the establishment of \textquotedblleft{}Nano
Functional Materials Technology Centre\textquotedblright{} (Grant:
SRNM/NAT/02-2005). J. M. thanks Mr. Ajit Jana for his help in DFT calculations. Institute HPCE facility was used for the computations. J. M. wants to thank UGC for SRF grant.\end{acknowledgments}

\bibliography{Paper}

\end{document}